\documentclass[
aip,
reprint,
amsmath,
amssymb,
hyphens
]{revtex4-1}
\usepackage{graphicx} 
\usepackage{dcolumn} 
\usepackage{bm}  
\usepackage[detect-all]{siunitx} 
\usepackage{doi}  
\usepackage{url}  
\providecommand{\urlprefix}{URL: } 
\hypersetup{colorlinks, linkcolor=blue, urlcolor=blue}  
\usepackage{xfrac}  
\usepackage{enumitem}  
\usepackage{pgfplots}  
\pgfplotsset{compat=newest}  
\usepackage[utf8]{inputenc}  
\DeclareUnicodeCharacter{2212}{-}
\usepackage{multirow}  
\newcolumntype{C}{>{\centering\arraybackslash}X}  
\newcolumntype{d}[1]{D{,}{\; \rightarrow \;}{#1}} 
\usepackage{adjustbox}  
\usepackage{stackengine}  
\usepackage{chngcntr}  

\DeclareSIUnit{\sq}{\ensuremath{\Box}}
\DeclareSIUnit{\wtpercent}{wt\%}

\begin{document}

\title{Design and Performance of Hafnium Optical and Near-IR Kinetic Inductance Detectors}

\author{Nicholas Zobrist}
\email[]{nzobrist@physics.ucsb.edu}
\author{Grégoire Coiffard}
\affiliation{\mbox{Department of Physics, University of California, Santa Barbara, CA 93106, USA}}
\author{Bruce Bumble}
\affiliation{\mbox{Jet Propulsion Laboratory, California Institute of Technology, Pasadena, California 91125, USA}}
\author{Noah Swimmer}
\author{Sarah Steiger}
\author{Miguel Daal}
\author{Giulia Collura}
\author{Alex B. Walter}
\author{Clint Bockstiegel}
\author{Neelay Fruitwala}
\author{Isabel Lipartito}
\author{Benjamin A. Mazin}
\homepage{http://www.mazinlab.org}
\affiliation{\mbox{Department of Physics, University of California, Santa Barbara, CA 93106, USA}}

\date{\today}
\begin{abstract}
We report on the design and performance of Microwave Kinetic Inductance Detectors (MKIDs) sensitive to single photons in the optical to near-infrared range using hafnium as the sensor material. Our test device had a superconducting transition temperature of \SI{395}{mK} and a room temperature normal state resistivity of \SI{97}{\micro \ohm \cm} with an $\mathrm{RRR} = \num{1.6}$. Resonators on the device displayed internal quality factors of around \num{200000}. Similar to the analysis of MKIDs made from other highly resistive superconductors, we find that modeling the temperature response of the detector requires an extra broadening parameter in the superconducting density of states. Finally, we show that this material and design is compatible with a full-array fabrication process which resulted in pixels with decay times of about \SI{40}{\micro \second} and resolving powers of \num{\sim 9} at \SI{800}{nm}.
\end{abstract}
\maketitle

Optical and near-IR (OIR) MKIDs are superconducting sensors capable of measuring the arrival time and energy of optical to near-infrared photons.~\cite{Mazin2013} They are less sensitive to false counts and radiation damage~\cite{Karatsu2016} than semiconductor devices operating in the same wavelength range and can achieve higher readout speeds. Moreover, each MKID is a high quality factor resonator which allows for natural frequency domain multiplexing and distinguishes the technology from other superconducting detectors. These advantages make arrays of OIR MKIDs useful as astrophysics cameras focusing on time-domain astronomy~\cite{Strader2016a, Szypryt2014, Strader2013} and high contrast imaging.~\cite{Walter2019, Meeker2018, Cook2015} To date, commissioned instruments have used either non-stoichiometric titanium nitride or platinum silicide alloys as the photon-sensitive material in the resonators and have achieved resolving powers, $R = E/\Delta E$, of up to \num{8} at \SI{800}{nm}.~\cite{Szypryt2016} This resolving power has been shown to be limited equally by stationary noise, generated by two-level systems (TLS) in the device and amplifiers in the readout chain, as well as an intrinsic variance in the photon signal pulse height, likely caused by phonon escape from the superconductor to the substrate during the initial photon energy down-conversion.~\cite{Zobrist2019}

Typically, each resonator is patterned in a lumped element design since it provides a relatively large photon-sensitive inductor with roughly uniform sensitivity. A microlens array can then be placed on top of the device to focus the light onto the inductor, increasing the fill factor to about \SI{90}{\percent}. To use this resonator geometry while maintaining a high enough responsivity to detect single photons, the inductors must be made from a high surface impedance and low $T_c$ superconductor, like TiN$_\text{x}$ or PtSi$_\text{x}$. Each of these materials, however, has its disadvantages. TiN$_\text{x}$ films suffer from non-uniformities in the gap energy which cause resonant frequency placement issues in large arrays, and PtSi$_\text{x}$ films can be expensive because of the required platinum sputter target. Additionally, successful fabrication of these materials relies on tightly controlling the deposition rates between the two elements and, in the case of PtSi$_\text{x}$, the annealing temperature, which may require time-consuming readjustment when calibrations drift.

An ideal sensing material for an OIR MKID, then, is a high surface impedance, low $T_c$, elemental superconductor which is common, compatible with standard MKID fabrication steps, and whose material properties are easy to control. It is also desirable for the energy resolution of these devices to be less sensitive to phonon-related degradation. Moving to an even lower transition temperature may be helpful for this last consideration since lower $T_c$ films have smaller quasiparticle creation energies and, for a given sensor responsivity, can be made with thicker films. Both of these effects have been shown to suppress phonon escape fluctuations.~\cite{Kozorezov2008} While these particular aspects are not further probed in this paper, they provide an important motivation for investigating lower $T_c$, elemental superconductors as alternative resonator materials for these detectors.

The lowest $T_c$ that can be used in an MKID is determined by the minimum achievable temperature of the employed refrigeration technology since operating the resonators above a temperature of about $T_c / 8$ can introduce excess readout noise and reduce the device's internal quality factor, $Q_i$.~\cite{DeVisser2011} Current instruments use adiabatic demagnetization refrigerators because of their relatively low cost, small size, temperature stability, and \SI{>10}{\hour} hold times. However, the standard base temperature of about \SI{100}{mK} limits the $T_c$ of an MKID in one of these systems to around \SI{800}{mK}, near the $T_c$ of TiN$_\text{x}$ and PtSi$_\text{x}$. Dilution refrigerators, comparatively, can reach temperatures of \SI{10}{mK} and lower. This system enables the use of transition temperatures down to \SI{100}{mK} as long as the readout frequencies are kept below twice the gap frequency to avoid significant unwanted quasiparticle generation.~\cite{Guruswamy2014a}

\begin{figure*}
    \includegraphics[width=\linewidth]{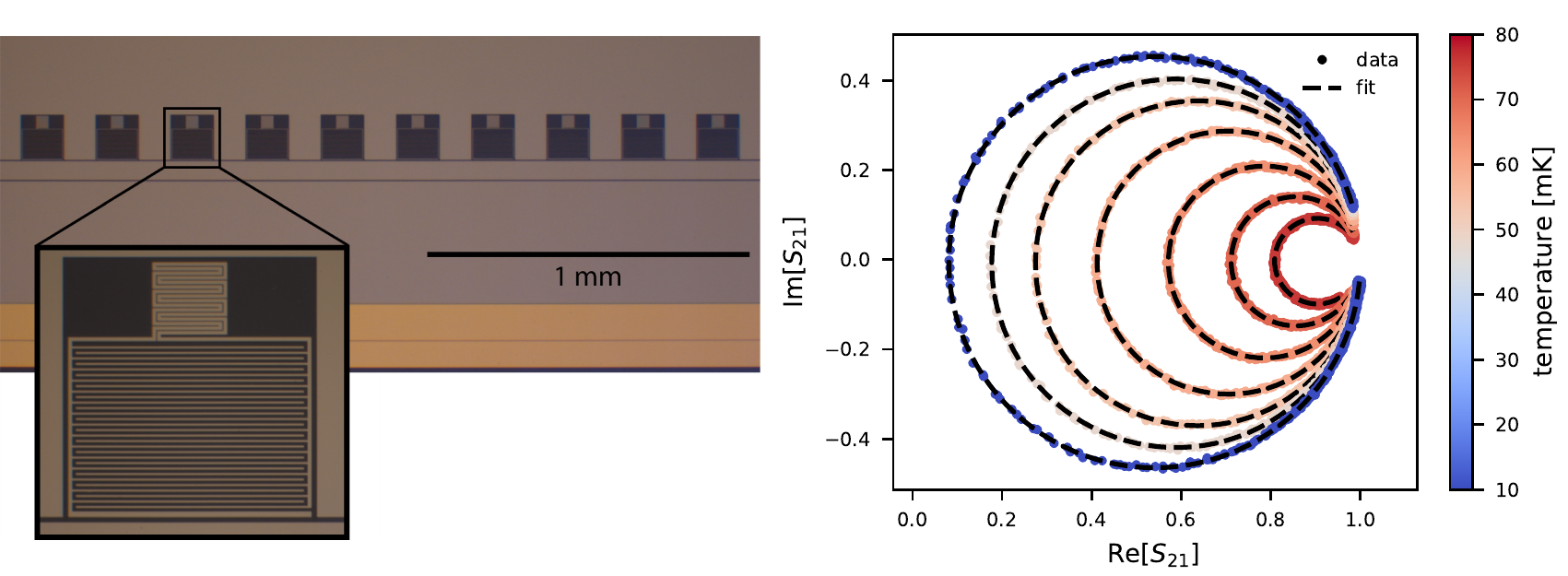}
    \caption{The test device is shown on the left. Dark grey, light grey, and yellow areas correspond to the sapphire substrate, hafnium film, and gold bond pad respectively.
    On the right, the normalized complex transmission of one of these resonators is shown for various temperatures. The fitted gain variations, phase offset, and cable delay are removed from the data in the plots for clarity. This resonator had a coupling and maximum internal quality factor of \num{17800} and \num{203000} respectively.}
    \label{fig:device}
\end{figure*}

There are several elemental superconductors with transition temperatures below \SI{800}{mK}. Here, we investigate hafnium as a potential candidate for this application. Its superconducting properties have been thoroughly investigated for use in superconducting tunnel junction detectors~\cite{Peacock1999, Kraft1998, Kozorezov2000} and transition edge sensors.~\cite{Hunacek2018, Lita2010} Sputtered films have a hexagonal close-packed (hcp) crystal structure and are in the local superconducting limit with critical temperatures ranging from \SIrange{140}{450}{mK} depending on deposition conditions. Furthermore, the surface impedance and quasiparticle lifetime of Hf films are similar to those in PtSi$_\text{x}$ and TiN$_\text{x}$ devices which mean that very little optimization needs to be done to transition detector array designs to this material.

A simple test device, shown in Figure~\ref{fig:device}, with ten resonators was used to measure the resonator quality and material properties of a \SI{125}{nm} thick hafnium film without exposing it to extra processing steps. The resonators are placed at a \SI{236}{\micro m} pitch, and each has a \SI{40 x 40}{\micro m} inductor and a \SI{100 x 130}{\micro m} capacitor---a similar design to that used in our current full-scale instruments.~\cite{Szypryt2017b} Gold wire bonds connect gold pads on the perimeter of the \SI{2.5 x 16.5}{mm} chip to its light-tight device box to improve thermalization. 

The detectors were designed to have resonance frequencies equally separated in a \SI{500}{MHz} bandwidth centered around \SI{5}{GHz} by tuning the capacitor leg lengths slightly between resonators.~\cite{[{The E\&M software used for the design can be found at }]sonnet} Assuming $L_s = \SI{16.7}{pH\per\sq}$ results in the closest agreement between the measured and simulated resonance frequencies, so we conclude this to be the device surface inductance. This value for the surface inductance corresponds to a kinetic inductance fraction, $\alpha$, of about \num{0.96} for these resonators. In addition, the superconducting transition temperature of the device was measured to be $T_c = \SI{395 \pm 5}{mK}$, where the error is dominated by the thermometer calibration uncertainty. More details about the fabrication of this device, experimental setup, and $T_c$ measurement can be found in the appendices.

Non-uniformity in $T_c$ across a wafer is a significant problem for TiN$_\text{x}$ MKIDs in large arrays because it yields resonant frequencies different from the design specifications. This hinders frequency multiplexing by depressing the pixel yield due to overlapping resonance frequencies. The film uniformity, therefore, is an important figure of merit for resonator materials used in large arrays. Both PtSi$_\text{x}$~\cite{Szypryt2016} and Ti/TiN multilayers~\cite{Vissers2013} have been shown to improve on this problem, so it is important to check that hafnium resonators have similar uniformity. In comparison, the standard deviation of the fractional error between the measured and designed frequencies on the test device was \num{1.2e-3} which is similar to the measured \num{7.1e-4} for PtSi$_\text{x}$ and compares well to the \num{1.1e-2} accuracy in TiN$_\text{x}$ resonators.~\footnote{The data for PtSi$_\text{x}$ and TiN$_\text{x}$ are taken from Ref.~\onlinecite{Szypryt2016} in which a line of 9 similar, lumped element resonators was used for each material at a pitch of \SI{444}{\micro m}. All of the resonators had kinetic inductance fractions near 1.}

The sheet resistance at room temperature can be used to predict $T_c$ uniformity across the whole wafer. We deposited a hafnium film using the process described above, and the sheet resistance was measured at 103 positions on the substrate using a non-contact Lehighton Lei 1510E-SA sheet resistance probe. The average sheet resistance over the \SI{100}{mm} diameter wafer was \SI[per-symbol=/, per-mode=symbol]{8.11 \pm 0.26}{\ohm\per\sq}, corresponding to a resistivity of \SI{97}{\micro\ohm\cm}. We measured a residual resistivity ratio (RRR) of \num{1.6} between room temperature and \SI{4.2}{K} with no additional change at temperatures approaching the superconducting transition. The \SI{3}{\percent} wafer uniformity is similar to that of PtSi$_\text{x}$ and much better than the measured \SI{\sim 20}{\percent} for TiN$_\text{x}$,~\cite{Szypryt2016, Vissers2013} matching the expectations set by the frequency placement accuracy in the test mask.

The resonance parameters were investigated by fitting the complex transmission coefficients to a four parameter resonance model~\cite{Khalil2012} with an additional five parameters to account for gain variations, a phase offset, and the cable delay.~\cite{Carter2017} Data was taken at temperatures from \SIrange{10}{80}{mK} and with a readout power well below the bifurcation power at about \SI{-120}{dBm}. The readout power is estimated from known attenuations, but an exact calibration does not exist for this system. No significant power dependence was found for $Q_i$ below the bifurcation power. At the lowest temperatures, the average $Q_i$ for these resonators saturates at about \num{200000}. An example of these fits for various temperatures is shown on the right side of Figure~\ref{fig:device}.

\begin{figure}
    \includegraphics[width=\linewidth]{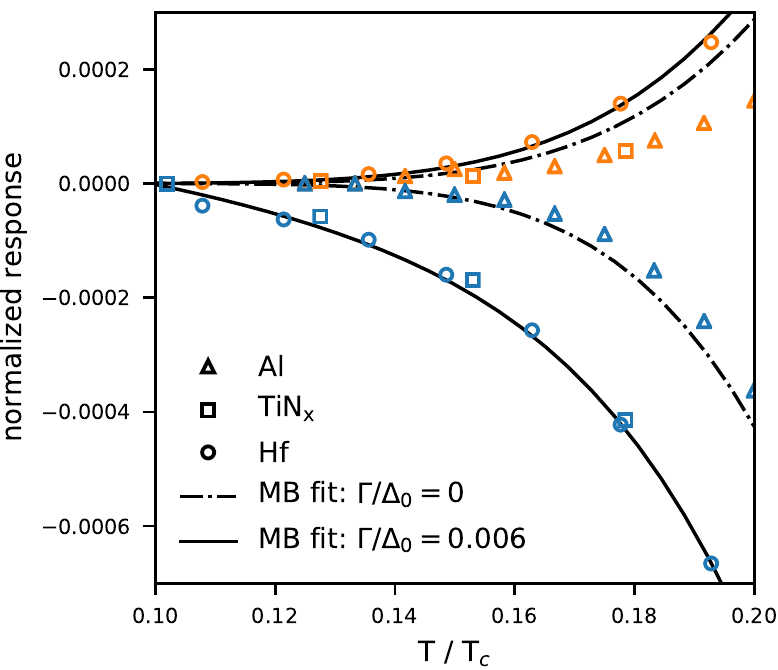}
    \caption{The normalized frequency response, $\sfrac{2 \delta f_r}{\left(f_r \alpha \gamma \right)}$ (blue), and dissipation response, $\sfrac{\delta Q_i^{-1}}{(\alpha \gamma)}$ (orange), for different materials are shown. The hafnium data is from the test device described in the text and is fit to the extended Mattis-Bardeen equations where the pair breaking parameter and gap energy were varied. The fitted responses with $\Gamma = 0$ are shown for comparison. Data for Al and TiN$_\text{x}$ are reproduced from J. Gao, M. R. Vissers, M. O. Sandberg, F. C. S. da Silva, S. W. Nam, D. P. Pappas, D. S. Wisbey, E. C. Langman, S. R. Meeker, B. A. Mazin, H. G. Leduc, J. Zmuidzinas, and K. D. Irwin, Appl. Phys. Lett. 101, 142602 (2012). with the permission of AIP Publishing.}
    \label{fig:response}
\end{figure}

The normalized frequency, $\sfrac{2 \delta f_r}{\left(f_r \alpha \gamma \right)}$, and dissipation, $\sfrac{\delta Q_i^{-1}}{(\alpha \gamma)}$, responses for the test device as a function of temperature are shown in Figure~\ref{fig:response} and compared to those for Al and TiN$_\text{x}$ resonators, where $\gamma$ is a film dependent constant ($\sfrac{1}{3}$ for Al and \num{1} for TiN$_\text{x}$ and Hf).~\cite{Zmuidzinas2012} Typically, the temperature responses are calculated via the complex conductivity using the Mattis Bardeen equations,~\cite{Mattis1958} but like resonators made from other superconductors with high disorder, the response of this device is not well described by this theory.

From the measured low temperature resistivity and a Fermi velocity of \SI{1.7e6}{m\per s},~\cite{Jepsen1975} we calculate a disorder parameter of $k_F l = 13$ using the Drude-Sommerfeld model. This value indicates similar but slightly less disorder than stoichiometric TiN and NbTiN.~\cite{Driessen2012} We therefore use a model for the complex conductivity that has been applied before to these materials.~\cite{Barends2008, Gao2012}  The Mattis-Bardeen equations are extended by adding a Lorentzian broadening parameter, $\Gamma$, to the superconducting density of states.~\cite{Herman2017} Physically, this parameter represents a Cooper-pair breaking scattering process caused by impurities in the film, which have the effect of modifying both the temperature dependence of the superconducting gap energy and the complex conductivity.~\cite{Herman2016} The extended Mattis-Bardeen equations fit well to the hafnium data when we also allow the zero temperature gap energy, $\Delta_0$, to vary. For hafnium, we find $\Gamma / \Delta_0 = 0.006$ and $\Delta_0 = 1.5 k_B T_c$. Other methods have been used to extend the Mattis-Bardeen equations to incorporate gap broadening and may be more or less applicable to hafnium depending on the exact broadening mechanism.~\cite{Noguchi2018, Zemlicka2015, Coumou2013} Thermal distribution functions are also assumed which do not capture potential non-equilibrium effects.~\cite{DeVisser2014a} As such, the fit should be interpreted to show a qualitative agreement with this model.

These results show promise for a practical MKID camera, but making a full array of detectors with around \num{2000} resonators per feedline requires additional fabrication steps than those used for the test device described above. A lower surface inductance superconductor, like niobium, is used for the transmission line to facilitate better impedance matching, and the ground planes on either side of the transmission line must be electrically connected with crossovers. Typically, we find the lowest TLS noise when the resonator metal is deposited first, ensuring a clean interface with the substrate. So, each of these extra steps has the potential to degrade the internal quality factors of the resonators. Unsurprisingly, initial attempts at fabricating a full hafnium MKID array produced unusable resonators with low $Q_i$. 

\begin{figure}
    \includegraphics[width=\linewidth]{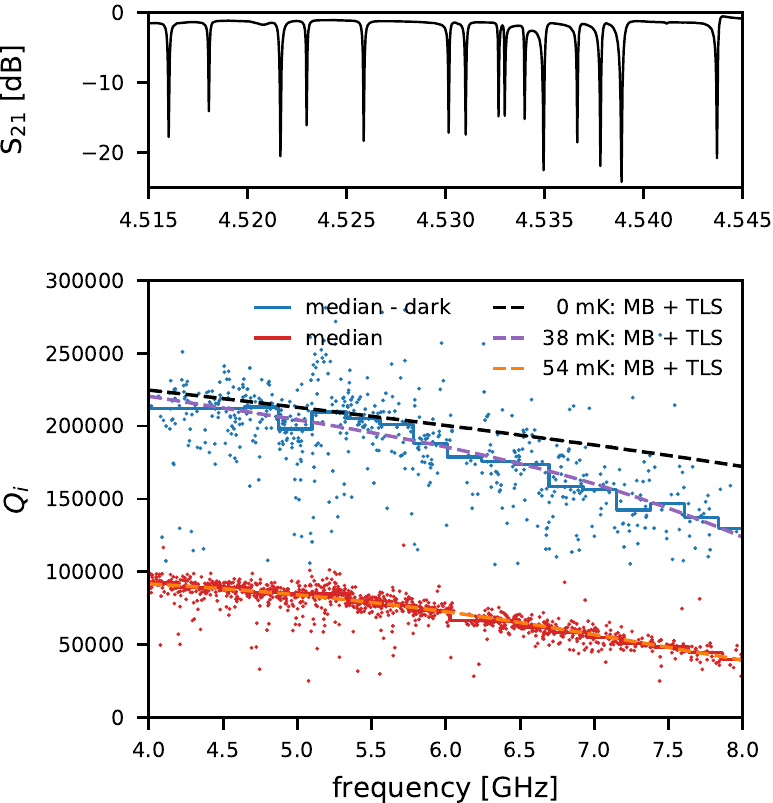}
    \caption{The top plot shows the magnitude of the forward scattering parameter over a \SI{30}{MHz} bandwidth around \SI{4.53}{GHz} for one feedline on a MEC-style hafnium array in a light-tight box. The designed resonant frequency spacing is \SI{2}{MHz}, and all of the \num{15} expected resonators showed up in this range. Below, the median $Q_i$ of the resonators on this array is shown as a function of binned frequency when the array is either in a light-tight box or exposed to the fridge radiation. The data points represent individual $Q_i$ measurements in each case. The dashed lines show a joint fit to the medians using the extended Mattis-Bardeen model with an additional two-level system component. The expected $Q_i$ dependence at zero temperature using this fit is plotted for comparison.}
    \label{fig:array}
\end{figure}

For PtSi$_\text{x}$, this problem was solved by adding a capping layer of tungsten to protect the resonators from the other fabrication steps and then removing it before testing.~\cite{Szypryt2017b} Tungsten was unavailable in our deposition system, so we used chromium as the hafnium capping layer to fabricate a \num{20440} pixel MEC-style array.~\cite{Szypryt2017b} Other potential capping materials were investigated, and more details can be found in the appendices. To adjust the film impedance to the MEC design, the hafnium film thickness was increased to \SI{200}{nm}. This change also had the effect of better matching the detector sensitivity to the desired \SIrange{800}{1400}{nm} wavelength range than in our test device. Figure~\ref{fig:array} shows a portion of the microwave transmission amplitude through the device as well as the measured distribution of internal quality factors at \SI{20}{mK}. We found a median $Q_i = \num{190000}$ across the array when in a light-tight box. Although, when the device box was fitted with a microlens array and exposed to the \SI{4}{K} radiation environment inside of the fridge, the median $Q_i$ dropped to \num{77000}. 

This behavior deviates from that seen in PtSi$_\text{x}$ arrays where the $Q_i$ does not change when exposed to the same radiation. We can understand this decrease by applying the extended Mattis-Bardeen model to the two data sets. Allowing the TLS loss, device temperature in each case, gap energy, and broadening parameter to vary, we find a joint fit with a TLS component corresponding to $Q_i \sim \num{270000}$, device temperatures of \SI{38}{mK} and \SI{54}{mK}, $\Gamma / \Delta_0 = 0.003$, and $\Delta_0 = 1.6 k_B T_c$. The TLS component represents an average, near-constant loss across this frequency and fitted-temperature range. 

In both cases, the fitted device temperature is hotter than the measured fridge temperature. The MEC-style arrays are much larger than the test device (\SI{24.6 x 22.5}{mm}), and while gold heat sinking is done along the border of the chip, it is possible that this extra area prevents the chip from fully cooling to the fridge temperature. Alternatively, the elevated effective device temperature may also be explained by a non-equilibrium phonon distribution in the superconductor caused by the background radiation.~\cite{Parker1975} It is unclear which mechanism is occurring, but the \SI{54}{mK} temperature suggests why this effect was not noticed in PtSi$_\text{x}$ arrays since they are operated near \SI{100}{mK}. An optical coating on the microlens array may be needed to block this radiation and avoid this effect in future hafnium devices.

\begin{figure}
    \includegraphics[width=\linewidth]{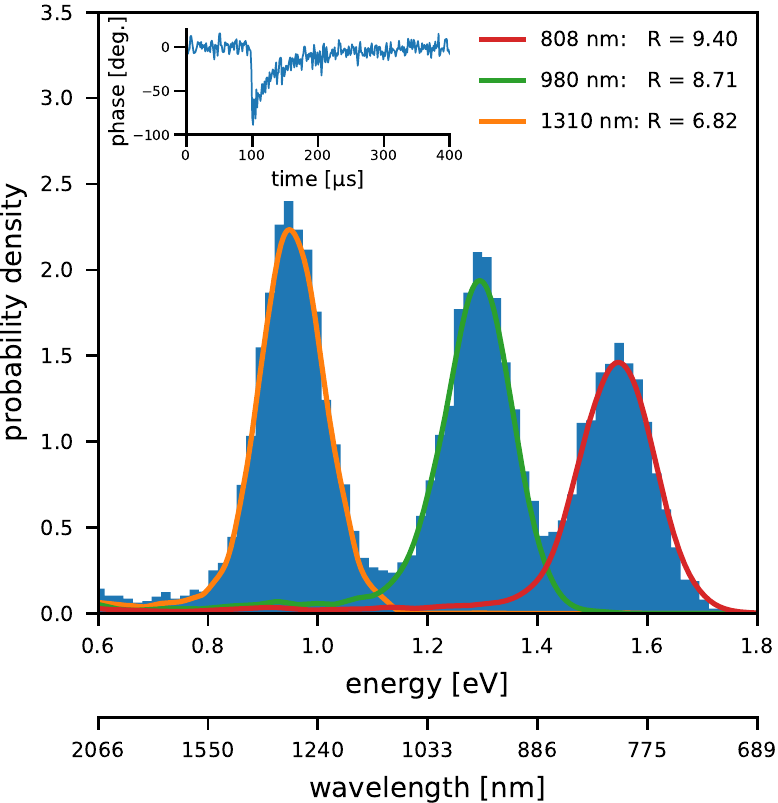}
    \caption{A composite spectrum for a resonator in the MEC-style array is plotted. The resonator was at a frequency of \SI{4.99942}{GHz}, had a $Q_i=\num{66100}$, $Q_c = \num{15700}$, and was measured at \SI{18}{mK}. The spectrum was generated by combining data collected with \SI{808}{nm}, \SI{980}{nm}, and \SI{1310}{nm} wavelength lasers individually. The solid lines correspond to kernel smoothed density estimations of the individual laser distributions and were used to compute the resolving power at each energy. In the inset, an example \SI{808}{nm} single photon event is shown. We measure average fractional frequency shifts of \num{3.2e-5} and \num{1.8e-5} for \SI{808}{nm} and \SI{1310}{nm} events respectively.}
    \label{fig:r}
\end{figure}

Although the $Q_i$ for the full array with the microlens is still significantly lower than in the test device, it is still high enough for this detector application. After probing a typical resonator at around \SI{-106}{dBm}, near the bifurcation power, we illuminated the array with three lasers spanning \SIrange{800}{1300}{nm} and measured photon absorption events. This data is shown in Figure~\ref{fig:r}, and we find that the resolving power of the hafnium MKID slightly exceeds that seen in TiN and PtSi$_\text{x}$ arrays.~\cite{Szypryt2017b,Mazin2013} For this device and readout, the resolving power is limited by noise coming primarily from two-level systems at low frequencies and from the HEMT amplifier at higher frequencies.~\cite{[{Specifications for the amplifier used in this paper can be found at }]LNF} The pulse decay time in this configuration was \SI{44}{\micro s} and \SI{34}{\micro s} in the phase and dissipation responses respectively.

In conclusion, we have demonstrated hafnium as a potential material for use in OIR MKID arrays. The film has a high resistivity indicating disorder on the same magnitude as stoichiometric TiN and NbTiN films, and a low $T_c = \SI{395}{mK}$. Measurements of the hafnium MKID temperature response show that it has slightly higher low temperature dissipation than other materials. This effect is shown to be comparable to the two-level system dissipation in a full-scale array. Because the TLS loss is still very low, the device noise is similar to devices made from PtSi$_\text{x}$ and TiN$_\text{x}$, giving a resolving power of about \num{9} at \SI{800}{nm}. 

N.Z. was supported throughout this work by a NASA Space
Technology Research Fellowship. The MKID arrays used were
developed under NASA Grant No. NNX16AE98G. This research
was carried out in part at the Jet Propulsion Laboratory and
California Institute of Technology, under a contract with the
National Aeronautics and Space Administration.

\appendix
\section{Hafnium Deposition} \label{sec:fab}
The test device was fabricated on a \SI{100}{mm} diameter a-plane sapphire wafer in a load-locked ultra-high vacuum sputtering system with a typical base pressure of \SI{e-7}{Pa}. To maintain a temperature near \SI{23}{\degree C}, the sample was held stationary and backed by an aluminum heat sink. The sputter source was a \SI{150}{mm} diameter DC magnetron with a hafnium target from Kamis Inc. that has two major impurities: zirconium and oxygen at the \SI{<1}{\wtpercent} and \SI{0.034}{\wtpercent} level respectively. The film was sputtered at a distance of \SI{10}{cm}, and the source power was controlled at \SI{500}{W}, corresponding to a deposition rate of \SI{40}{nm\per min} with an argon pressure \SI{0.33}{Pa}. These conditions produce films with a high compressive stress of \SI{1.3}{GPa} for a \SI{125}{nm} thick film.

After deposition, the sample was patterned with a Canon FPA-3000 EX-3 deep-UV stepper with a numerical aperture of \num{0.65} using \SI{248}{nm} Cymer laser. GKR-6760 photoresist was used and spun on top of a \SI{100}{nm} layer of PMMA to assist in the removal of the photoresist after the process step. The PMMA layer was oxygen dry etched after developing the resist with AZ MIF-300. The hafnium film was then etched in an Unaxis Shuttleline ICP tool with a BCl$_3$ + \SI{40}{\percent} Cl$_2$ gas mixture at a pressure of \SI{1.33}{Pa}, with a \SI{400}{\watt} plasma power, and with a \SI{30}{\watt} bias power. This system produces vertical sidewall profiles with sub-micron sized features. The chlorine ICP step was followed by oxygen plasma ashing and a deionized water rinse prior to removing the photoresist in solvents. The oxygen plasma ashing was done at relatively high pressure and low power on a graphite electrode in a dedicated RIE system. Gold pads were added to the perimeter of the chip by lift-off e-beam evaporation. Gold wire bonds connecting these pads to the device box ensure good thermalization of the chip during testing. 
\section{Experimental Setup} \label{sec:setup}
All low temperature measurements of the hafnium films were taken in a Bluefors LH400 dilution refrigerator. Optical access is achieved through N-BK7 glass windows on the outer casing and the \SI{50}{K},  \SI{4}{K}, and still radiation shields. The windows at \SI{50}{K} and \SI{4}{K} are coated with an IR blocking filter that passes wavelengths between \SI{773}{nm} and \SI{1465}{nm}. There is no radiation shield on the mixing chamber or cold plate temperature stages.

The tested devices were coupled to the same microwave line in the refrigerator. Signals are transmitted to the detectors through a lossy CuNi coaxial cable inside of the fridge with an additional \SI{30}{dB} of attenuation at the mixing chamber stage. Up to six different devices can be connected at the mixing chamber stage at a time between two Radiall R573423600 microwave switches, modified to work at cryogenic temperatures. On the output of the second switch is a B320H DC block, which is included to prevent the switch operation from damaging the HEMT amplifier at the \SI{4}{K} stage. The low noise factory \SIrange{4}{8}{GHz} G3PO HEMT amplifier is connected to the DC block via a superconducting coax and a short SMA to G3PO adaptor cable.

Measurements of the test device transmission and MEC-style device optical data were taken with a standard homodyne readout scheme. The microwave signal is sent into the fridge with a Anritsu MG37022A signal generator. The output signal is mixed down with a Marki IQ0318L IQ mixer and then low pass filtered at \SI{400}{kHz} before being digitized by a National Instruments PCI-6120 DAQ at a \SI{0.8}{MHz} sample rate. For the \SIrange{4}{8}{GHz} transmission measurements on the MEC-style device, an Agilent Technologies E5071C vector network analyzer was used instead of the homodyne readout to improve the data acquisition speed.

\section{Superconducting Transition Temperature} \label{sec:tc}
The superconducting transition temperature of the test device was determined with a four-wire measurement using a hafnium sample from the same wafer as the test device. The sample was varnished to an OFHC copper mount on the mixing chamber stage of the refrigerator, and aluminum-silicon wire bonds were used to connect to copper traces on a FR-4 circuit board. Low thermal conductivity, \SI{127}{\micro m} diameter, twisted-pair, manganin wires with lengths \SI{>30}{cm} were used to connect the circuit board to a heat sink on the cold plate of the fridge, and a calibrated RX-102A-CD-0.05B thermometer from Lake Shore Cryogenics was bolted next to the sample. Thermometer and sample resistances were simultaneously measured by a Lake Shore 370AC resistance bridge with a model 3716 scanner, and the temperature was controlled using a resistive heater, ramping upward at a rate of \SI{1}{mK\per \minute} with the PID controller on the resistance bridge. 

\begin{figure}
    \centering
    \includegraphics[width=\linewidth]{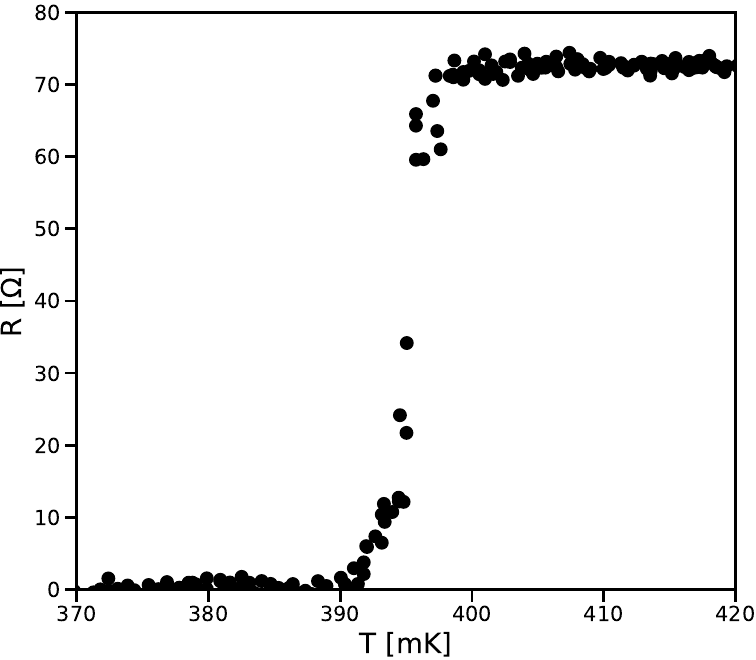}
    \caption{Plotted is the resistance of the \SI{125}{nm} thick hafnium $T_c$ sample as a function of temperature in the region of the superconducting transition.}
    \label{fig:rvst}
\end{figure}

The sample resistance versus temperature data are displayed in Figure~\ref{fig:rvst}, showing a clear transition from the superconducting to normal state over about \SI{5}{mK}. The scatter in the resistance measurement is primarily from the thermometer readout. Changing the measurement range to its smallest setting confirms that the resistance at \SI{380}{mK} is below at least \SI{100}{n\ohm}. 

\section{Capping Layer} \label{sec:cap}
The capping layer on an MKID array protects the resonator layer from the other fabrication processes and is removed as the final step. A good capping layer, therefore, must etch with the same chemistry as the resonator layer so that the resonator can be patterned, be removable with an etch that stops on the resonator layer, and not degrade the internal quality factor of the resonators. It is also preferable to deposit the capping layer in situ with the resonator layer to ensure a clean interface.

For hafnium, using the test device design, we investigated the effect of three different candidate capping layers, aluminum, niobium, and chromium on different portions of the same wafer. A section of the wafer was left uncovered as a control. Each of these materials etch with the same process used for hafnium, described in appendix~\ref{sec:fab}. After the device was fabricated, the aluminum was removed with a wet etch in MIF-300 resist developer for \SI{1}{\minute}; the niobium was removed with a CF$_4$ + \SI{10}{\percent} O$_2$ ICP etch at \SI{0.933}{Pa} and low power for \SI{7}{\minute}; and the chromium was removed with a wet etch with a standard chrome mask etchant produced by Cyantek. Both the niobium and chromium capping layers had no effect on the measured $Q_i$ when compared to the control device, but the aluminum capping layer reduced the $Q_i$ of the device by over \SI{50}{\percent}.

\bibliography{references}

\end{document}